\documentstyle{article}
\newcommand{\hochpunkt}[1]{\mbox{$^{\raisebox{.3ex}{\scriptsize #1}}_{\raisebox{.6ex}{\hspace{.17em}.}}$}}
\newcommand{\hoch}[1]{\mbox{$^{\raisebox{.3ex}{\scriptsize #1}}$}}
\begin{document}

\begin{center}
{\LARGE \bf Time resolved spectroscopy and photometry of three little known 
bright cataclysmic variables: LS~IV~-08$^{\rm o}$~3, HQ~Monocerotis and 
ST~Chamaeleontis}\footnote{Based 
on observations taken at the Observat\'orio do Pico dos Dias / LNA}

\vspace{1cm}

{\Large \bf Albert Bruch}

\vspace{0.5cm}
Laborat\'orio Nacional de Astrof\'{i}sica, Rua Estados Unidos, 154,
CEP 37504-364, Itajub\'a - MG, Brazil

\vspace{0.5cm}
{\Large \bf Marcos P.\ Diaz}

\vspace{0.5cm}
Instituto de Astronomia, Geof\'{\i}sica e Ci\^encias
Atmosf\'ericas, Universidade de S\~ao Paulo, Rua do Mat\~ao, 1226,
CEP 05508-090, S\~ao Paulo, Brazil

\vspace{0.8cm}

(Text published in: New Astronomy, Vol.\ 50, p.\ 109 -- 119 (2017))
\vspace{0.8cm}

{\bf Abstract}

\vspace{0.5cm}
\end{center}

\begin{quote}
{\parindent0em As part of a project to better characterize comparatively 
bright but so far
little studied cataclysmic variables in the southern hemisphere, we have 
obtained spectroscopic and photometric data of the nova-like variables
LS~IV~-08$^{\rm o}$~3 and HQ~Mon, and of the Z~Cam type dwarf nova ST~Cha.
The spectra of all systems are as expected for their respective types.
We derive improved orbital ephemeris of LS~IV~-08$^{\rm o}$~3 and map its
accretion disk in the light of the H$\alpha$ emission using Doppler
tomography. We find that the emission has a two component origin, arising
in the outer parts of the accretion disk and possibly on the illuminated
face of the secondary star. The light curve of LS~IV~-08$^{\rm o}$~3 exhibits
a low level of flickering and indications for a modulation on the orbital 
period. Spectroscopy of HQ~Mon suggests an orbital period of $\approx 
5\hochpunkt{h}15$ which is incompatible with previous (uncertain) 
estimates. The light curves show the typical low scale flickering of 
UX~UMa type nova-like systems, superposed upon variations on longer time
scales. During one night a modulation with a period of $\approx 41\hoch{m}$
is observed, visible for at least 4 hours. However, it does not repeat
itself in other nights. A spectroscopic orbital period of 
$\approx 5\hochpunkt{h}5$ is derived for ST~Cha. A previously suspected
period of 6\hochpunkt{h}8 (or alternatively 9\hochpunkt{h}6), based on
historical photographic photometry is incompatible with the spectroscopic
period. Moreover, we show that our new as well as previous
photometry does not contain evidence for the quoted photometric period.}
\vspace{0.5cm}

{\parindent0em {\bf Keywords:}
Stars: novae, cataclysmic variables --
Stars: individual: LS~IV~-08$^{\rm o}$~3 --
Stars: individual: HQ~Mon -- 
Stars: individual: ST~Cha}
\end{quote}

\section{Introduction}
\label{Introduction}

Cataclysmic variables (CVs) are interactive binaries where a late type, 
low mass star which is normally on or close to the main sequence transfers
matter to a white dwarf. For a general and encompassing introduction to
CVs, see Warner (1995) or Hellier (2001).
The number of known systems of this kind has grown 
enormously in recent years mainly due to numerous detections of CVs in
large scale surveys. However, most of these newly discovered systems
are rather faint in their normal brightness state. Therefore, the
characterization of their individual properties is expensive because is 
requires large telescopes.

On the other hand, it is much easier to perform detailed studies of the
brighter CVs, most of which are known for a long time. It is therefore
surprising that even among these stars an appreciable number has not
yet been adequately characterized to be certain about basic parameters
such as the orbital period. In some cases even the very class membership
is not confirmed. 

We therefore started a small observing program aimed at a better understanding
of these so far neglected stars. For this purpose we selected a number of
little studied southern CVs and suspected CVs bright enough   
($m_{\rm vis} \le 15\hoch{m}$) to be easily observed with comparatively small
telescopes. The emphasis of this program lies on photometry with a high
time resolution aimed at the detection of short and medium time scale 
variations such as flickering and orbital variability. However, in a few 
cases the photometric observations could be complemented by spectroscopic 
time series. The first results of this program, concerning the dwarf nova 
MU~Cen, were recently published by Bruch (2016).

Here, we report on three stars for which spectra were taken in addition to 
light curves. These are the nova-like variables LS~IV~-08\hoch{o}~3 and
HQ~Mon, and the Z~Cam type dwarf nova ST~Cha. Nova-like variables are
CVs with accretions disks in a bright state which do not show outbursts
such as the dwarf novae (or, alternatively, may be considered to be in
permanent outburst). Sometimes, they are subdivided into UX~UMa stars
which always remain on approximately the same brightness level, and VY~Scl
stars which occasionally assume a low state at much fainter magnitude than
normal. The Z~Cam stars are systems which, instead of alternating 
between quiescent and outburst states due to a limit cycle instability in 
their accretion disks as is usual with dwarf novae, sometimes
get stuck for some time at a brightness just below the outburst level. This
phenomenon in known as ``standstill''.

HQ~Mon and ST~Cha entered the
observing program mainly because no time resolved photometry was ever 
published and because the orbital periods as quoted in the most recent 
on-line edition of the Ritter \& Kolb catalogue (Ritter \& Kolb 2003) 
are uncertain 
and require confirmation. On the other hand, LS~IV~-08\hoch{o}~3 was quite 
thoroughly characterized by Stark et al.\ (2008). It entered the observing 
program only as a backup target. However, as it turns out, the present 
additional observations nicely complement those of Stark et al.\ (2008).

In Sect.~\ref{Observations and data reductions} we report briefly on the
observations and data reductions. The results for the indivivual objects 
are then dealt with in turn in Sects.~\ref{LS IV -8 3} -- \ref{ST Cha}. We
summarize the most important conclusions in Sect.~\ref{Conclusions}.

\section{Observations and data reductions}
\label{Observations and data reductions}

All observations were obtained at the
Observat\'orio do Pico dos Dias (OPD), operated by the Laborat\'orio Nacional
de Astrof\'{\i}sica, Brazil. For photometry the 0.6-m Zeiss and the
0.6-m Boller \& Chivens telescopes were used. Spectroscopy was performed using
the 1.6-m Perkin Elmer telescope.
  
Time series imaging of the field around the target stars was performed
using cameras of type Andor iXon EMCCD DU-888E-C00-\#BV
and iKon-L936-B equipped with back illuminated, visually optimized CCDs.
In order to resolve the expected rapid flickering variations the integration
times were kept short. Together with the small readout times of the detectors
this resulted in a time resolution of the order of 5\hoch{s}. In order to 
maximize the count rates in these short time intervals no filters were
used. Therefore, it was not possible to calibrate the stellar magnitudes.
Instead, the brightness was expressed as the magnitude difference between the 
target and a nearby comparison star. This is not a severe limitation in view 
of the purpose to the observations. 
A summary of the photometric observations is 
given in Table~\ref{Journal of photometric observations}. Not all of them
were useful. Parts with an elevated noise level due to passing clouds will not
be regarded here. 

\begin{table*}

\caption{Journal of photometric observations}
\label{Journal of photometric observations}

\hspace{1ex}

\begin{tabular}{llcccc}
\hline
Name      & Date & Start & End    & Time  & Number  \\
          &      & (UT)  & (UT)   & Res.  & of      \\
          &      &       &        & (s)   & Integr. \\
\hline
LS~IV~-08\hoch{o}~3  & 2015 May 20 & \phantom{2}1:10 & \phantom{2}6:16
          & \phantom{1}6\phantom{.5}  &           3\,449 \\
          & 2015 May 21 & \phantom{2}3:35 & \phantom{2}6:52
          & \phantom{1}6\phantom{.5}  &           1\,971 \\ [1ex]
HQ Mon    & 2014 Mar 26 & 23:19           & 23:52
          & \phantom{1}5.5            & \phantom{1\,}346 \\
          & 2014 Mar 28/29 & 23:05 & \phantom{2}1:52
          & \phantom{1}5.5            & 2\,094           \\ 
          & 2014 Nov 21 & \phantom{2}3:20 & \phantom{2}7:34
          & \phantom{1}5\phantom{.5}  & 2\,923           \\ 
          & 2015 Feb 13 & \phantom{2}0:59 & \phantom{2}5:02
          & \phantom{1}5\phantom{.5}  & 2\,645           \\ [1ex] 
ST Cha    & 2014 Mar 29 & \phantom{2}2:52 & \phantom{2}5:22 
          & \phantom{1}5.5            & 1\,466           \\
          & 2014 Apr 30 & 21:38 & 22:00 
          & \phantom{1}5\phantom{.5}  & \phantom{1\,}256 \\ 
          & 2014 Jun 17/18 & 21:25       & \phantom{2}0:18 
          & \phantom{1}5\phantom{.5}  & 1\,800           \\ [1ex]
\hline
\end{tabular}
\end{table*}
%
The spectra were taken in 2015, February and March, 
using the Boller \& Chivens spectrograph of OPD. 
Details are given in Table~\ref{Journal of spectroscopic observations}.
An Andor iKon-L936-BR-DD camera was employed. Integration times were 
15\hoch{m} throughout. Exposures of a He-Ar lamp for wavelength calibration
were taken after every second stellar exposure. From the FWHM of the lines in
the comparison spectra a spectral resolution of $\approx$4~\AA\ is
estimated. The spectral range of the March data encompassed H$\alpha$
and H$\beta$. However, due to a error of the instrumental configuration, 
H$\alpha$ was not included in the February spectra. 
The observing conditions were 
quite variable, ranging from photometric to periods heavily affected by clouds.
Therefore, some of the spectra remained severely underexposed and are only of
limited usefulness. In view of these difficulties no attempt was made to 
flux calibrate the data. 

\begin{table*}
\caption{Journal of spectroscopic observations}
\label{Journal of spectroscopic observations}

\hspace{1ex}

\begin{tabular}{llcccc}
\hline
Name      & Date & Start & End    & Number  & Spectral \\
          &      & (UT)  & (UT)   & of      & Coverage \\
          &      &       &        & Spectra & (\AA)    \\
\hline
LS~IV~-08\hoch{o}~3  & 2015 Mar 24    & \phantom{2}5:33 & \phantom{2}8:35 
          & 11          & 4704 -- 6796   \\ 
          & 2015 Mar 25    & \phantom{2}4:35 & \phantom{2}8:21
          & 12          & 4705 -- 6797   \\
          & 2015 Mar 26    & \phantom{2}4:24 & \phantom{2}6:00
          & \phantom{1}6& 4706 -- 6798   \\
          & 2015 Mar 27    & \phantom{2}4:06 & \phantom{2}8:27
          & 16          & 4707 -- 6799   \\
          & 2015 Mar 28    & \phantom{2}4:34 & \phantom{2}7:31
          & \phantom{1}9& 4709 -- 6801   \\ [1ex]
HQ Mon    & 2015 Fev 13    & \phantom{2}0:57 & \phantom{2}4:46 
          & 12          & 4342 -- 6448   \\
          & 2015 Fev 14    & \phantom{2}2:48 & \phantom{2}4:31 
          & \phantom{1}6& 4343 -- 6449   \\
          & 2015 Mar 24    & 21:51           & 22:07 
          & \phantom{1}1& 4708 -- 6800   \\
          & 2015 Mar 25    & 21:39           & 22:50 
          & \phantom{1}2& 4709 -- 6801   \\
          & 2015 Mar 26/27 & 21:37           & \phantom{2}1:56 
          & 15          & 4708 -- 6800   \\ [1ex]
ST Cha    & 2015 Fev 13    & \phantom{2}4:54 & \phantom{2}7:11 
          & \phantom{1}4& 4339 -- 6445   \\
          & 2015 Mar 23/24 & 23:54           & \phantom{2}5:22 
          & 14          & 4702 -- 6794   \\ 
          & 2015 Mar 24/25 & 22:40           & \phantom{2}3:30 
          & 16          & 4702 -- 6794   \\ 
          & 2015 Mar 25/26 & 22:58           & \phantom{2}4:14 
          & 18          & 4704 -- 6798   \\ 
          & 2015 Mar 27    & \phantom{2}2:01 & \phantom{2}3:56 
          & \phantom{1}6& 4707 -- 6799   \\ [1ex]
\hline
\end{tabular}
\end{table*}
%

Basic reductions of the data (bias subtraction, flat-fielding, spectral 
extraction) were performed with IRFF. All further data reduction and analysis
was done using the MIRA software system (Bruch 1993), with the exception
of the Doppler mapping (Sect.~\ref{LS IV -08 3 Spectroscopy}) for which an
IRAF/SPP user code was employed.

Aperture photometry was applied to the images of the target stars, the
primary comparison stars and several check stars in the field. The time
axis of the resulting light curves is expressed in UT throughout this paper,
unless stated otherwise. In contrast, the times of mid-exposure of the spectra,
used to measure the orbital motion of the target stars, were transformed into
barycentric Julian Dates, following Eastman et al.\ (2010).
In order to remove the instrumental response curve
from the uncalibrated spectra, they were normalized to the continuum. The
raw spectra contained strong absorption features in particular in the 
wavelength ranges 
$\lambda \lambda$ 5860 -- 6020~\AA\ (mostly just to the red of the Na~D lines),
$\lambda \lambda$ 6260 -- 6340~\AA\ and
$\lambda \lambda$ 6450 -- 6620~\AA\ (to the blue and around H$\alpha$).
They can be attributed to atmospheric absorptions. Therefore, the normalized
spectra were divided by a suitably scaled normalized atmospheric transmission
spectrum (degraded to match the observed spectral resolution). This 
procedure was optimized for the strong absorptions close to Na~D and around 
H$\alpha$, at the expense of an imperfect elimination of the absorptions in
the third range, resulted in spurious structures centred on
$\lambda$ $\sim$6280~\AA\ in LS~IV~-08\hoch{o}~3 and HQ~Mon. With
few exceptions which will be appropriately specified below radial velocities
of lines in individual spectra were measured relative to the
corresponding lines in the average spectrum by means of cross correlation:
the maximum of a Gaussian fit to 
the cross-correlation function was used to calculate the radial 
velocity difference between the average and the individual spectra.
Throughout this paper we define as spectroscopic phase zero the 
negative-to-positive zero-crossing of the H$\alpha$ emission line radial 
velocity curve. In the ideal case (radially symmetric emission distribution 
confined to the accretion disk) this corresponds to the upper conjunction of 
the secondary star with respect to the white dwarf. 

\section{LS~IV~-08$^{\rm o}$~3}
\label{LS IV -8 3}

We will start the discussion of the results with the best known
of the three target stars, LS~IV~-08$^{\rm o}$~3. At 
$V \approx 11\hochpunkt{m}5$ (H{\o}g et al. 2000) it is also the brightest.
LS~IV~-08$^{\rm o}$~3 was originally classified as 
an OB star (Nassau \& Stephensen, 1963). Stark et al.\ (2008) narrate more 
of the history of our knowledge about this object. They performed a detailed 
spectroscopic study and found that LS~IV~-08$^{\rm o}$~3 
is a binary with a period of 0.1952894 days.
The characteristics of the system suggested a reclassification
as a nova-like variable of the UX~UMa subtype. Stark et al.\ (2008) also
present some time resolved photometry which reveals low scale (a few
hundredths of a magnitude) apparently stochastic variations superposed on
slight orbital variations.

\subsection{Spectroscopy}
\label{LS IV -08 3 Spectroscopy}

\begin{figure}
   \parbox[]{0.1cm}{\epsfxsize=14cm\epsfbox{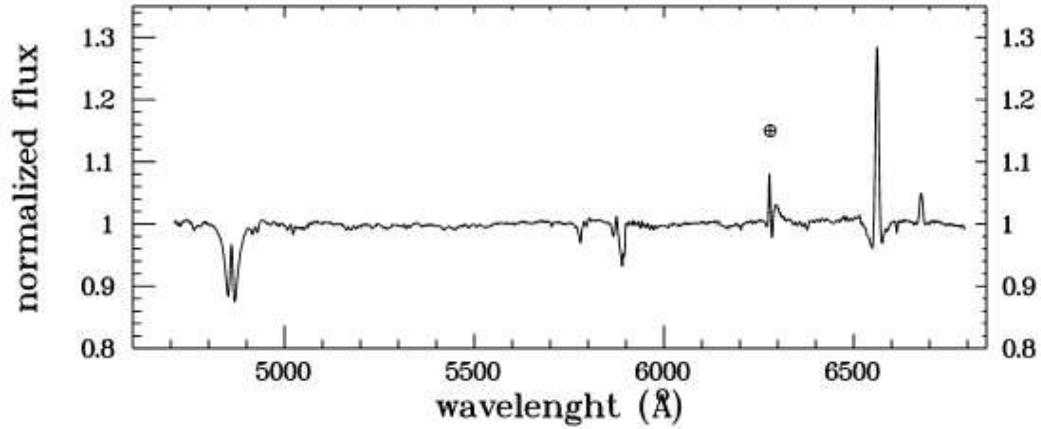}}
      \caption[]{Average continuum normalized spectrum of LS~IV~-08\hoch{o}~3.
                 (The marked structure is an artifact caused by an imperfect
                 removal of atmospheric absorption features.)} 
\label{ls-iv-avspec}
\end{figure}

Spectra of LS~IV~-08\hoch{o}~3 were obtained in five nights in 2015, March.
The average spectrum, weighting the individual spectra by their mean count 
rates, is shown in Fig.~\ref{ls-iv-avspec}. 
The marked structure around $\lambda$ $\sim$6280~\AA\ resulted from an 
imperfect elimination of atmospheric absorption lines (see
Sect.~\ref{Observations and data reductions}).

\input epsf

The spectrum resembles very much that shown in Fig.~2 of Stark et al.\ (2008)
and which is quite typical for CVs with accretion disks in the bright state: 
Balmer emission components decreasing in strength from H$\alpha$ to H$\beta$
are superposed upon broad absorption troughs which are stronger for the higher
Balmer lines. It appears that the centre of the H$\alpha$ absorption is 
somewhat shifted to the blue with respect to the emission component. This is
not the case for H$\beta$.
He~I $\lambda \lambda$ 5876~\AA\, and 6678~\AA\, are also seen
in emission\footnote{The impression that He~I $\lambda$~5876~\AA\, is
superposed on the blue flank of the broad Na~D absorption may have been
enhanced by uncertainties in the continuum normalization in this range.}.
The Na~D absorption at $\lambda \lambda$ 5890-5896~\AA\, is not expected in
the spectrum of a nova-like variable and is interpreted as interstellar by 
Stark et al.\ (2008). This is confirmed here (see below). Finally, the
absorption feature at $\lambda$~5780~\AA\, is a diffuse interstellar band.
Not surprisingly, we measure its equivalent width to be identical to that 
measured by Stark et al.\ (2008).  

The strength of the H$\alpha$ emission line and the good S/N-ratio of the
spectra permitted to apply the double Gaussian 
convolution method proposed by Schneider \& Young (1980) to measure the
line position. The separation $a$ of the two Gaussians was chosen to be
between 1~\AA\, and  7~\AA, fixing their width to 1~\AA. We then used the 
diagnostic diagrams first introduced by Shafter (1983) and later 
refined by Shafter et al.\ (1986) in order to determine the optimal orbital 
solution. The period was fixed to the value found by Stark et al.\ (2008).

\begin{figure}
   \parbox[]{0.1cm}{\epsfxsize=14cm\epsfbox{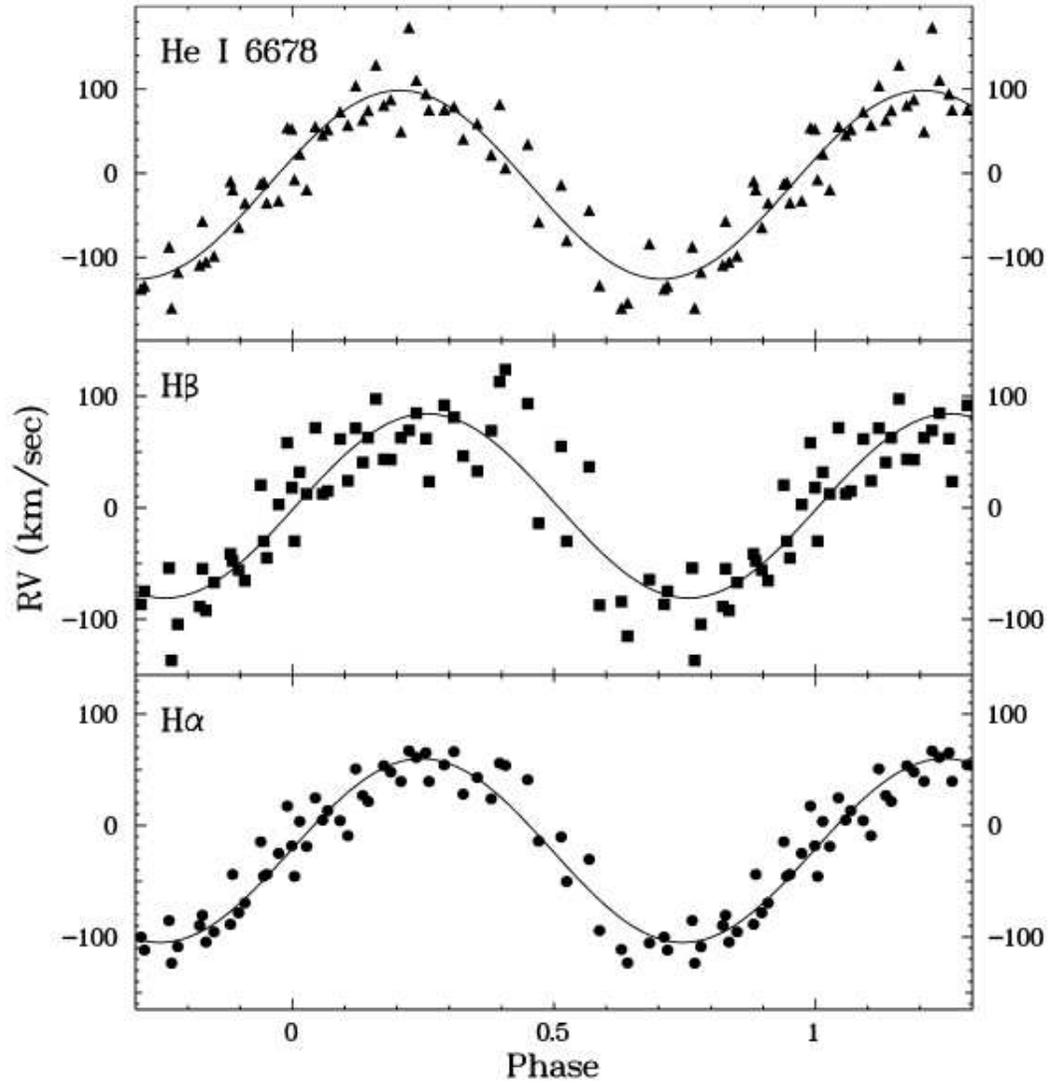}}
      \caption[]{Radial velocity curve of the He~I $\lambda$~6678~\AA\ (top), 
                 H$\beta$ (centre) and H$\alpha$ (bottom)
                 emission lines of LS~IV~-08\hoch{o}~3. The zero points of the
                 He~I and H$\beta$ curves are arbitrary, while for H$\alpha$ 
                 heliocentric velocities are plotted.} 
\label{ls-iv-rv}
\end{figure}

It was found that the dependence of the orbital solution on $a$ is quite 
weak. Within the sampled range of $a$, the amplitude of the radial velocity (RV)
variations did not vary by more than 6~km/sec, the $\gamma$-velocity remained
stable within 7~km/sec, and the phase shift was limited to 0.012. Thus, there
is no strong indication that different parts of the line profile follow 
distinct radial velocity curves (but see the results of Doppler mapping of
H$\alpha$ that will be presented below). In order to be definite, and 
following the
criterium normally used in the interpretation of the diagnostic diagrams, we 
choose the solution for $a = 5.2$~\AA\, as reference. This is the values at 
which the relative error $\sigma_K/K$ of the RV amplitude assumes a (shallow) 
minimum. The corresponding RV curve, after applying the heliocentric 
correction, is shown in the lower frame of Fig.~\ref{ls-iv-rv} together with 
the best fit sine curve.

The sine fit resulted in an amplitude of the RV curve of $K = 82 \pm 4$~km/sec.
This is slightly higher than the values measured by Stark et al.\ (2008) 
($77 \pm 4$)\, but still compatible within the error limits. However, the
presently found $\gamma$ velocity ($-21 \pm 3$~km/sec) is rather different
from their value ($-44 \pm 3$~km/sec). 

The weaker H$\beta$ emission did not warrant the application of the double
Gaussian convolution method in order to determine the line position. Therefore,
the cross correlation method was applied in three different ways: (i) using 
the entire line profile 
(absorption and emission component), (ii) masking the emission component and
cross-correlating only the absorption component, and (iii) cross-correlating
only the emission component, after dividing the entire line profile by a 
high order polynomial fit to the absorption component (emission masked). 
The resulting radial velocity values were very similar. In the central
frame of Fig.~\ref{ls-iv-rv} the RV curve, folded on the orbital period, of
the emission component is shown together with the best fit sine curve.
At $K = 83 \pm 7$~km/sec, the amplitude is almost identical to the one
measured for H$\alpha$. This is another indication that their are no strong
differences in the contribution to the hydrogen emission lines from different 
parts of the binary system. Due to the cross-correlation method applied here, 
the velocity zero-point of H$\beta$ is, of course, arbitrary.

The orbital period $P_{\rm orb}$
of LS~IV~-08\hoch{o}~3 derived by Stark et al.\ (2008) is
based on spectra taken between 2004 and 2007. The present new data enlarge
the total time base and thus enable a refinement of the period. The
accuracy of the value quoted by Stark et al.\ (2008) is high enough to
preserve cycle counts up to the present epoch. They used the same phase
convention as we did (see their Fig.~3). However, adopting their
ephemeris, we observe a phase difference of $0.17 \pm 0.09$ (H$\alpha$) and
$0.18 \pm 0.09$ (H$\beta$) with respect to the radial velocities of 2015, March.
Assigning (arbitrarily) twice as much weight to the phase difference of 
H$\alpha$ than to that of H$\beta$, and attributing it to a slight period 
error, results in refined ephemeris:
\begin{displaymath}
{\rm BJD} = 2457107.7527 (19) + 0.1952913 (3) \times E
\end{displaymath}
where $E$ is the cycle number. The epoch is close to the mean epoch of the
present observations. The errors are given in units of the last decimal digits.
For the epoch, it corresponds simply to the phase difference found between
the best sine fit to the H$\alpha$ and H$\beta$ RV curves. The error of the
period is estimated from the error quoted by Stark et al.\ (2008), assuming
that it scales inversely with the total time base of the observations. 

The He~I $\lambda$~6678~\AA\, emission line behaves slightly different from
the Balmer lines. Measuring the line position via cross-correlation with the
mean line profile yields the radial velocity curve, folded on $P_{\rm orb}$
using the above ephemeris, shown in upper frame of Fig.~\ref{ls-iv-rv}.
With $112 \pm 6$~km/sec the best fit sine curve has an amplitude which is
significantly larger than that of the Balmer lines. Such differences in the
RV amplitudes of different lines are not uncommon in cataclysmic variables
(see, e.g., the dramatic difference of the RV amplitude 
of H$\beta$ and He~II $\lambda$ 4686 \AA\, in HR~Del; Bruch 1982)
and can complicate a lot dynamical interpretations of line motions. Moreover,
there is a definite phase shift of $-0.05$ between the He line and the Balmer 
lines. Both findings point at differences in the emission sites of helium and
hydrogen. We refrain from trying to measure the radial velocity of He~I
$\lambda$~5876~\AA\, because of its superposition upon the blue flank of the
Na~D absorption.

Finally, for the sake of exercise, we also cross-correlated the Na~I~D 
lines of the individual spectra with their average, carefully
avoiding the He~I $\lambda$~5786~\AA\, emission line located on the blue
flanc of the absorption feature. As expected, no radial velocity variations
were found, confirming the identification by Stark et al.\ (2008) of these 
lines as interstellar.

\begin{figure}
   \parbox[]{0.1cm}{\epsfxsize=14cm\epsfbox{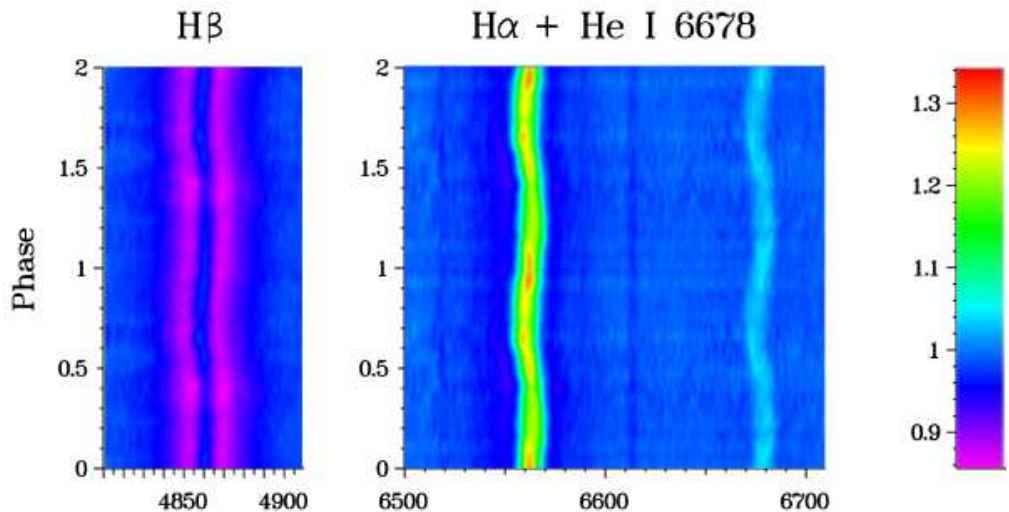}}
      \caption[]{Spectra of H$\beta$ and H$\alpha$ + He~I
                 $\lambda$~6678~\AA\, as a function of orbital phase. 
                 The scale of the colour bar at the right is given
                 in units of the continuum intensity.}
\label{ls-iv-2d}
\end{figure}

Fig.~\ref{ls-iv-2d} shows the trailed spectra of H$\beta$ and H$\alpha$ + He~I
$\lambda$~6678~\AA\, as a function of phase. The radial velocity variations
of all lines are readily visible. As mentioned earlier, the H$\alpha$ emission
is not centralized within the shallow absorption trough (see the asymmetry of
the dark blue shade on both sides of the emission in the figure). In their
high-resolution HET spectra of LS~IV~-08\hoch{o}~3 Stark et al.\ (2008)
detected changes in the shape of the H$\alpha$ emission profile which they
consider to be composed of a broad and a narrow component moving in anti-phase.
They tentatively identify the former as arising from the accretion disk and 
the latter from the side of the  mass donating star facing and being
illuminated by the primary component of the CV.
 
\begin{figure}
   \parbox[]{0.1cm}{\epsfxsize=14cm\epsfbox{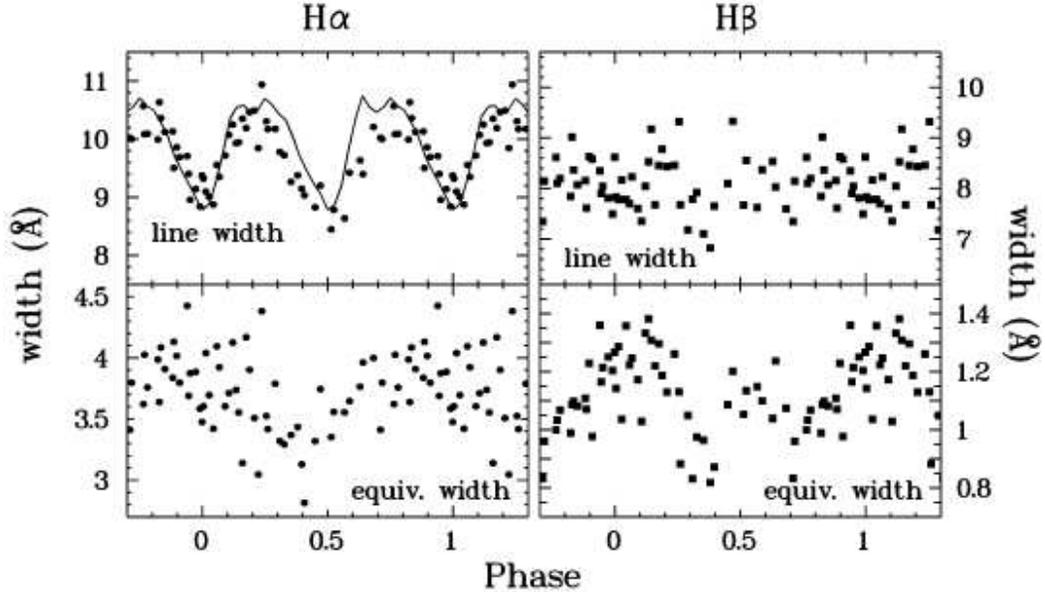}}
      \caption[]{Line width (FWHM of a 
                 Gaussian fitted to the line) and equivalent width of the 
                 emission components of H$\alpha$ and H$\beta$ as a function 
                 of phase. The solid line in the upper left frame represents
                 the line width reconstructed from the Doppler map (see text).} 
\label{ls-iv-lineprops}
\end{figure}

At the much lower resolution of the present data these details cannot be 
seen directly. However, a close inspection of Fig.~\ref{ls-iv-2d} shows that
the normalized strength of the emission components of both, H$\alpha$ and 
H$\beta$, exhibits variations as a function of phase. This is also seen in 
Fig.~\ref{ls-iv-lineprops} where in the lower frames the equivalent width of 
the emission (expressed as positive values for simplicity) is plotted as a
function of phase. In spite of considerable noise, a clear phase dependence
is visible in both lines. We also measured the line width (upper frames of
Fig.~\ref{ls-iv-lineprops}, expressed as the FWHM of a Gaussian
fitted to the emission. While no systematics can be seen in the case of 
H$\beta$, the H$\alpha$ line width exhibits a strikingly clear modulation
on half the orbital period, assuming minima at phase 0 and 0.5.

In order to verify if this modulation can be explained by structures in the
accretion disk, a Doppler map of the H$\alpha$ emission was calculated
using an algorithm known as filtered back projection. The mathematics related
to this method are described in detail by Rosenfeld \& Kak (1982). Before 
processing the data, the continuum around H$\alpha$ and the average absorption
profile, approximated (and interpolated beneath the emission) by a high order
polynomial, were subtracted from the individual spectra. The orientation of
the Doppler map depends on the orbital phase. Having no independent reliable
information (e.g., through eclipse measurements), we trust that the 
spectroscopic zero point of the phase given by the orbital ephemeris refers 
to the upper conjunction of the secondary star.

\begin{figure}
   \parbox[]{0.1cm}{\epsfxsize=14cm\epsfbox{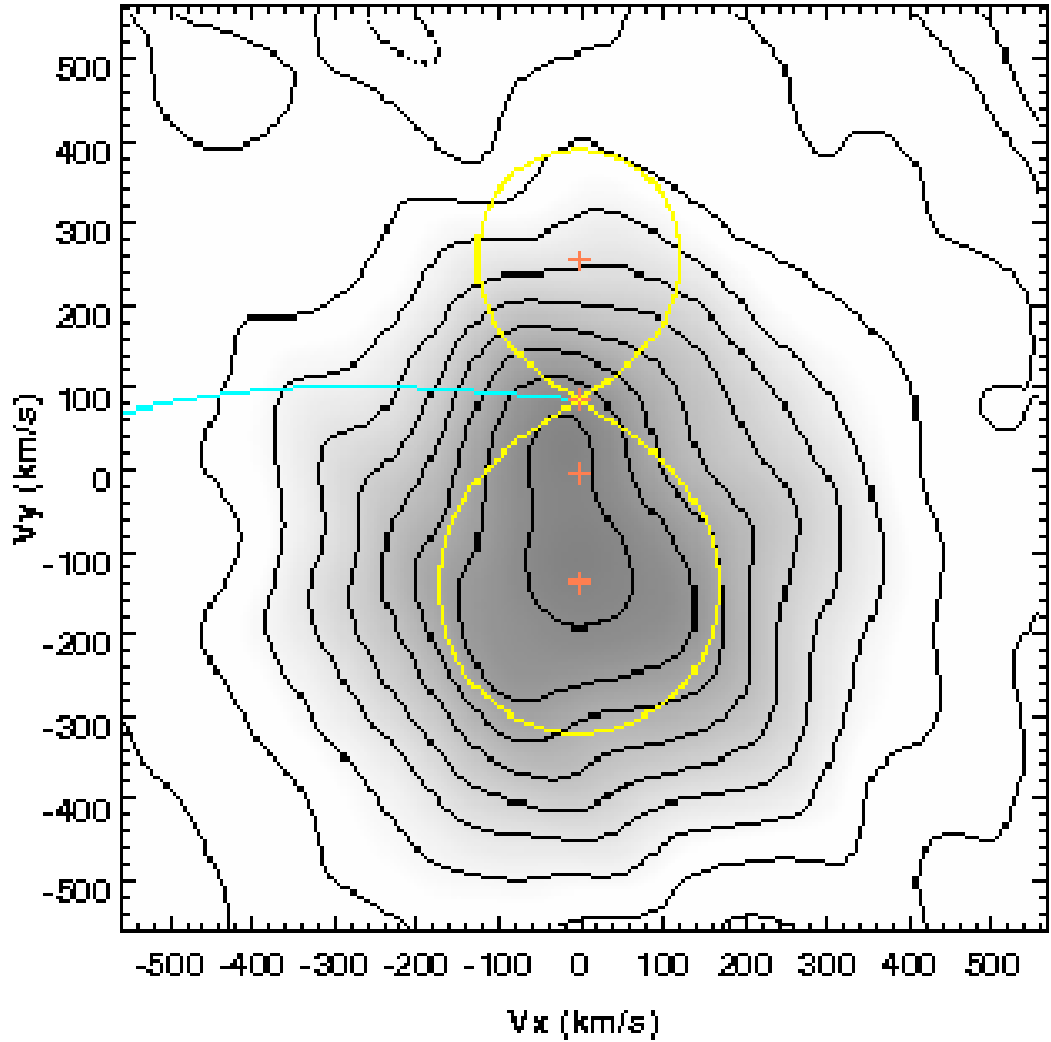}}
      \caption[]{Doppler map of the H$\alpha$ emission line of 
                 LS~IV~-08\hoch{o}~3, expressed on a linear flux scale.
                 The yellow contours indicate the location of the Roche
                 lobe, assuming (arbitrarily) an orbital inclination of
                 60\hoch{o}, and masses of 0.75~$M_\odot$ and 0.43~$M_\odot$, 
                 respectively, for the primary and secondary components of 
                 system. The blue line then indicates the expected stream
                 trajectory from the mass donor star.} 
\label{ls-iv-dopmap}
\end{figure}

The resulting Doppler map is shown in Fig.~\ref{ls-iv-dopmap}. Only for
illustrative purposes we include in the map as a yellow line the Roche lobe 
contours of the
binary, assuming -- in the absence of any more detailed information about the
respective parameters -- arbitrarily an orbital inclination of 60\hoch{o},
a primary star mass of 0.83~$M_\odot$, i.e.\ the mean mass of the white dwarf
in CVs according to Zorotovic et al.\ (2011), and a secondary star mass of
0.43~$M_\odot$, based on the semi-empirical relation between the 
mass of the secondary stars in CVs and their orbital period
derived by Knigge et al.\ (2011). Thus, the assumed mass ratio is 
$q = M_2/M_1 = 0.52$. The blue line is then the trajectory of the
mass transfer stream from the secondary to the accretion disk.
The resolution (FWHM) of the map is 
$\sim$200~km/s, corresponding to 4.4~\AA, comparable to the spectral resolution
quoted in Sect.\ \ref{Observations and data reductions}.  

The limited resolution of the Doppler map does not permit to identify many
details. At high velocities the map shows symmetric emission around the
white dwarf, resembling the Doppler maps of bright steady state disks with
single peaked emission components (e.g., the SW~Sex type
system V347~Pup; Diaz \& Hubeny 1999). There is no indication of a hot
spot which, if present, should be placed at elevated velocities on the stream
trajectory (blue line). 
Most of the emission is confined to low velocities, representing
the outer regions of the accretion disk. 
The emission structure is somewhat elongated along the line connecting
the stellar components of LS~IV~-08\hoch{o}~3. On close inspection, two 
separated nuclei can be distinguished, one of them centred on the primary star.
The other one is shifted into the direction of the secondary star. It
would be close to the inner Lagrangian point if 
$q\,$\mbox{$^{\raisebox{-.9ex}{$>$}}_{\raisebox{.6ex}{$\sim$}}$} 0.6 is chosen.
Considering the standard deviation of CV white dwarf masses around their
average ($0.23\ M_\odot$, Zorotovic et al.\ 2011) such a mass ratio can, of
course, easily be realized. Moreover, in combination with the assumed secondary
star mass it does not violate stability criteria derived from recent binary
population synthesis models (Schreiber et al. 2016). 
The origin of this emission
may therefore be on the illuminated surface of the secondary star.
This is then compatible with the interpretation of the narrow emission
component identified by Stark et al.\ (2008) as arising on the mass donor.  

The elongated structure is responsible for the modulation of the width of
the H$\alpha$ emission seen in Fig.~\ref{ls-iv-lineprops}. Whenever the line of
sight is aligned with the line connecting the two nuclei their movement is 
perpendicular to the line of sight and the velocity dispersion attains a 
minimum. A quarter of an orbit later the nuclei move along the line of sight to 
the observer and thus the velocity dispersion is maximal. This can also be seen
when reconstructing the emission line from the Doppler map. For this purpose 
the map was collapsed in the direction of the line of sight as a function
of the orbital phase. The FWHM of a Gaussian fit to the resulting distribution
is superposed upon the directly measured line width as a solid line in the 
upper left frame of Fig.~\ref{ls-iv-lineprops}. 
It shows a similar modulation as the directly observed line width. 

\subsection{Photometry}
\label{LS IV -08 3 Photometry}

\begin{figure}
   \parbox[]{0.1cm}{\epsfxsize=14cm\epsfbox{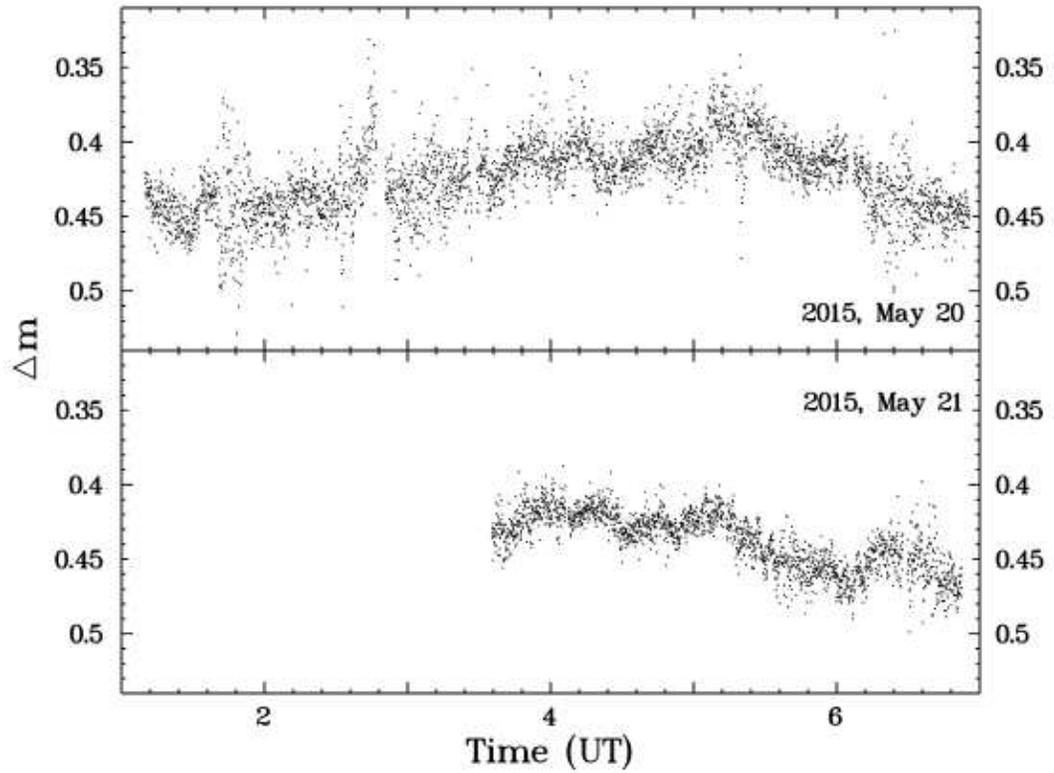}}
      \caption[]{Differential light curves of LS~IV~-08\hoch{o}~3 in
                 two nights in 2015, May.}
\label{ls-iv-lightc}
\end{figure}

Light curves of LS~IV~-08\hoch{o}~3 were observed in two consecutive nights 
(2015, May 20 and 21). They are shown in Figs.~\ref{ls-iv-lightc}.
Differential magnitudes are given with 
respect to the comparison star TYC 5642-0042-1 
($V_T = 11\hochpunkt{m}18$; $B_T = 12\hochpunkt{m}39$). This is the same
star used by Stark et al.\ (2008) for their photometry of LS~IV~-08\hoch{o}~3.
Both light curves are characterized by gradual variations on hourly time scales,
superposed by irregular flickering activity with a typical amplitude of the
order of 0\hochpunkt{m}06 (average of the two light curves, ignoring variations
on time scales longer than $\approx$1\hoch{h} and occasional stronger flares.
This compares well with the flickering amplitudes of other UX~UMa type novalike
variables. Using results of Beckemper (1995) we calculate these to
be typically 0\hochpunkt{m}05 (median of 63 light curves of 7 
objects). The corresponding numbers for other CV types are (with increasing
median amplitude): dwarf novae in outburst (rise, maximum or decline;
excluding supermaxima ocurring in short period dwarf novae of SU~UMa type): 
0\hochpunkt{m}07 (122 light curves of 28 objects); classical novae in 
quiescence: 0\hochpunkt{m}10 (80 light curves of 10 objects); VY~Scl stars in 
their high state: 0\hochpunkt{m}18 (34 light curves of 5 objects) and dwarf 
novae in quiescence: 0\hochpunkt{m}29 (79 light curves of 15 
objects)\footnote{These numbers should only be regarded as rough 
indications.
A systematic study on this and other statistical properties of the
flickering in many CVs is being prepared by the first author.}.

\begin{figure}
   \parbox[]{0.1cm}{\epsfxsize=14cm\epsfbox{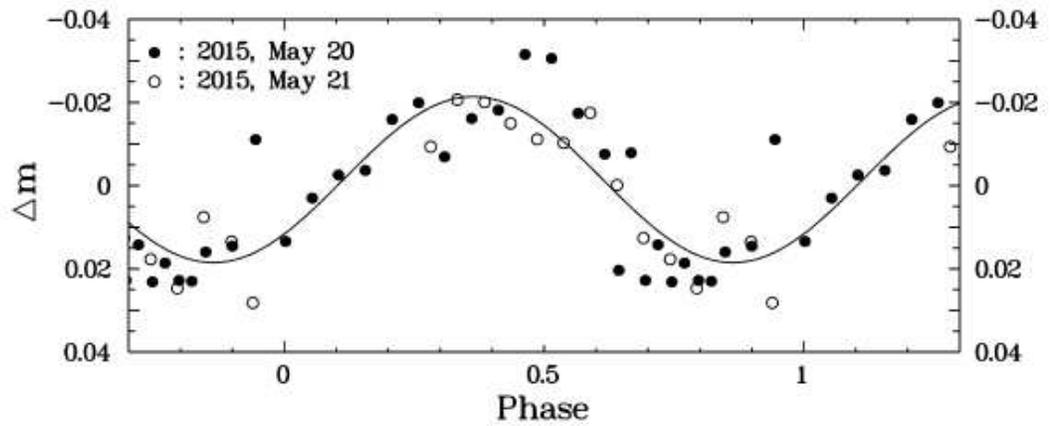}}
      \caption[]{Light curves of LS~IV~-08\hoch{o}~3 folded on the 
                 spectroscopic period. Data from different nights are
                 distinguished by different symbols.}
\label{ls-iv-folded}
\end{figure}

Although the data consist only of two light curves the finding of 
Stark et al.\ (2008) of low amplitude orbital modulations of the brightness
of  LS~IV~-08\hoch{o}~3 can be substantiated by the present observations.
To show this, first the average nightly magnitude was subtracted from the
light curves in order to eliminate possible night-to-night variations
common in CVs and also seen in LS~IV~-08\hoch{o}~3 
(see Fig.~6 of Stark et al.\ 2008). They were then binned into intervals 
of $0.01 P_{\rm orb}$ and time was transformed into barycentric Julian
Date. Folding the resulting data on $P_{\rm orb}$ then results in the
light curve shown in Fig.~\ref{ls-iv-folded} where data from different nights
are distinguished by different symbols.

The phase folded light curve exhibits a sinusoidal modulation with a total
amplitude of $\approx$0\hochpunkt{m}04. Some deviating points can be 
explained as being caused by particular flickering spikes. The brightness
assumes a maximum close to spectroscopic phase 0.36 and a minimum near 
phase 0.86. While the current observations thus confirm the presence of
orbital variations, their phasing is just the opposite of that found by
Stark et al.\ (2008) (see the lower frame of their Fig.~6).

\section{HQ Monocerotis}
\label{HQ Monocerotis}

HQ~Mon is classified as RV: (RV~Tau star with questionable classification) 
in the on-line version of the General Catalogue of Variables Stars (Samus 
et al.\ 1975), i.e.\ a radially pulsating supergiant. It was discovered by 
Morgenroth (1933) who describes its variations as ``slowly variable'' 
with a range between 13\hochpunkt{m}5 and 14\hochpunkt{m}5. The RV~Tau 
classification first appears in the study of van Bueren (1950).
Lloyd Evans (1984) found the spectrum to contain shallow hydrogen lines,
and says that the $UBV$ colours ``confirm that it is a B star'' According
to the tables of Schmidt-Kaler (1982) this is true 
for the measured mean $U-B=-0.77$ but not for his $B-V=0.04$ which is too
red. Instead, the colours are compatible with those of a cataclysmic
variable (Bruch \& Engel 1994).

Van Bueren's (1950) 
classification of HQ~Mon is questioned by Wahlgren et al.\ (1985)
who describe the spectrum as containing hydrogen emission components within 
broad absorption lines together with emission of He~II $\lambda$ 4686. 
Wahlgren (1992) notes that the star resembles a cataclysmic variable
``with a period of 32 days'' (which is evidently not meant to be the orbital 
period). A low resolution spectum is reproduced by Zwitter \& Munari (1994). 
It has a continuum rising steeply to short wavelengths. H$\alpha$ and He~II 
$\lambda$4686 are in emission, but only a broad absorption is seen at H$\beta$. 
The Na~D absorption lines are also discernible. The orbital period of 
7\hochpunkt{h}59 quoted in the Ritter \& Kolb catalogue is based on an 
informal communication by J.\ Patterson 
\footnote{http://cbastro.org/communications/news/messages/0301.html} and 
requires confirmation. 

\subsection{Spectroscopy}
\label{HQ Mon Spectroscopy}

\begin{figure}
   \parbox[]{0.1cm}{\epsfxsize=14cm\epsfbox{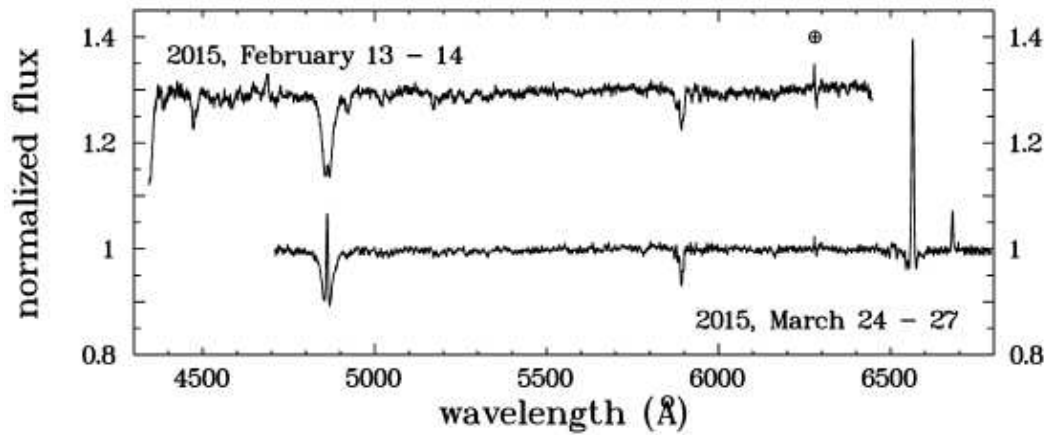}}
      \caption[]{Average continuum-normalized spectra of HQ~Mon
                 in 2015, February and March. The February spectrum has been
                 shifted upward by 0.3 units for clarity. (The marked 
                 structure is an artifact caused by an imperfect
                 removal of atmospheric absorption features.)}
\label{hqmon-avspec}
\end{figure}

Spectra of HQ~Mon were obtained in two nights in 2015, February and in three
nights in 2015, March (see Table~\ref{Journal of spectroscopic observations}). 
The average spectrum, weighing the individual spectra by their mean count rates,
for the February and March observing runs are shown in 
Fig.~\ref{hqmon-avspec} (for clarity the February spectrum was shifted 
upwards by 0.3 units). 

The March spectrum resembles very much that of LS~IV~-08\hoch{o}~3 but the
H$\beta$ emission component and (less so) the He~I $\lambda$~6678~\AA\ line
are stronger. Again, the Na~D lines are probably interstellar. The 
spectrum reproduced by Zwitter \& Munari (1994) differs somewhat from the 
March spectrum in the sense that the emission component of H$\beta$ is absent. 
But in this respect it is similar to the present February spectrum, where only 
a faint emission core appears at the bottom of the broad H$\beta$ absorption 
line. Thus, significant variations of the hydrogen emission line strength in 
HQ~Mon occur. Unfortunately, H$\alpha$ was not covered in February, but in 
compensation the respective spectrum confirms the presence of He~II 
$\lambda$~4686~\AA\ in emission.

The absorption line at $\lambda$ $\sim$4473~\AA\ 
was also seen by Wahlgren et al.\ (1985) who attribute it to
He~I $\lambda$ 4471~\AA\ and Mg~II $\lambda$ 4481~\AA. They contemplate a
classification of HQ~Mon as a Be star, in line with the presence of this
absorption, but reject this idea because emission of He~II $\lambda$ 4686~\AA\
would not be expected in a Be star. He~I $\lambda$ 4471~\AA\, is seen
in emission in numerous cataclysmic variables (most often in dwarf novae
in quiescent) but relevant compilations of CV spectra 
(Bruch, 1989; Bruch \& Schimpke, 1992; Zwitter \& Munari, 1994, 1995, 1996,
Munari et al., 1997)
also contain examples where the line appears in absorption, usually in
systems with accretion disks in the bright state

Radial velocities of the H$\alpha$ emission line (March data) were measured, 
using the cross correlation method. The same was done for the combined 
absorption and emission line of H$\beta$ (February and March data). 
In view of the strong difference of the strength of the emission component
during the two epochs the individual line profiles were cross-correlated 
not with the overall mean profile, but with the mean profile of 
the respective month. Alternatively, the emission component was masked and
the cross-correlation was performed with the average profile of all data. 
Both approaches yielded consistent results. Hereafter, we will regard the
radial velocities derived from the entire line profile.

\begin{figure}
   \parbox[]{0.1cm}{\epsfxsize=14cm\epsfbox{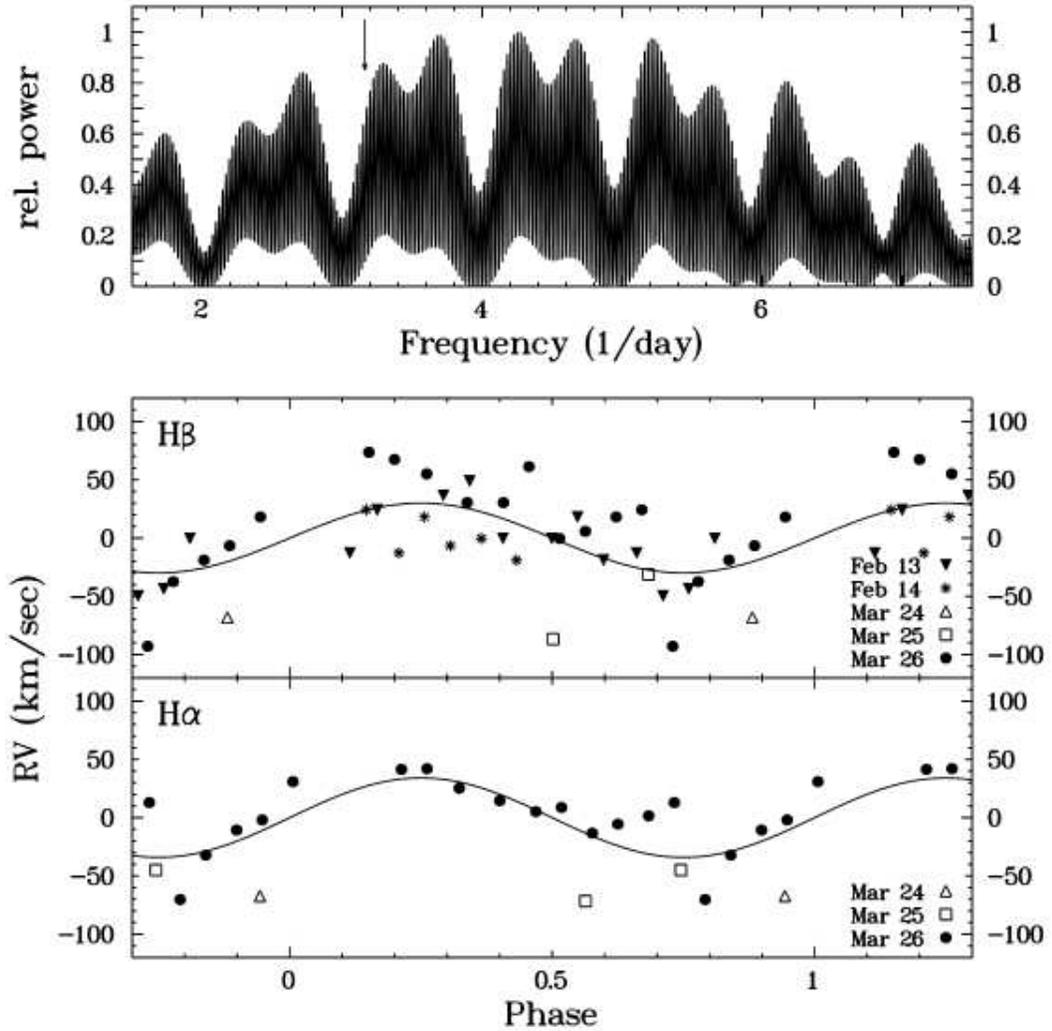}}
      \caption[]{{\it Top:} Normalized power spectrum of the H$\beta$ radial
                 velocities of HQ~Mon. The arrow indicates the frequency
                 corresponding to the orbital period informally proposed
                 by J.\ Patterson. {\it Bottom:} Radial velocities of H$\beta$
                 (upper frame) and H$\alpha$ (lower frame) folded on the
                 period $P_{\rm orb} = 0\hochpunkt{h}21462$ suggested by the
                 present data (see text).} 
\label{hqmon-rv}
\end{figure}

The H$\alpha$ radial velocities alone do not permit a reliable determination
of a period because most of the March observations were performed in a single
night and only one and two, respectively, spectra were taken in the two
preceding nights. The H$\beta$ data provide a longer time base, including
the February observations. But even so their temporal distribution is
far from ideal for a period analysis. The upper frame of Fig.~\ref{hqmon-rv}
contains the Deeming (1975) power spectrum of all H$\beta$ radial 
velocities. It contains a complicated alias pattern due to the extremely 
unequal data spacing. None of the peaks stands out to suggest itself as 
corresponding to the only reasonable radial velocity period. The four highest 
peaks suggest periods of 
0\hochpunkt{d}192,
0\hochpunkt{d}215,
0\hochpunkt{d}234, and 
0\hochpunkt{d}270, respectively.
We will concentrate subsequently on the second of these values because using
this choice results in an acceptable\footnote{or should we say: the
least unacceptable?} (although still quite noisy)
radial velocity curve not only for H$\beta$ (shown in the second frame of 
Fig.~\ref{hqmon-rv} together with the best fit sine curve) when the data
are folded on that period, but also for H$\alpha$ (lower frame of the figure). 
Adopting any of the alternative periods results in even less convincing radial
velocity curves. Using the best fit sine curves RV amplitudes were measured to 
be $34 \pm 7$~km/sec for H$\alpha$ and $30 \pm 5$~km/sec for H$\beta$,
respectively.

Thus, the radial velocity measurements suggest an orbital period of 
$P_{\rm orb} = 0\hochpunkt{d}21462 \pm 0\hochpunkt{d}00025$ 
($\approx 5\hoch{h} 9\hoch{m}$) which, however, we only regard as
preliminary. Here, the error was estimated from the dispersion of
a Gaussian fit to the power spectrum peak correponding to $P_{\rm orb}$. 
But note that the fine structure of the alias pattern caused by the separation
of $\sim$40 days between the February and March observations permits other
choices separated by 0.024 cycles/day from the orbital frequency corresponding
to $P_{\rm orb}$. Moreover, we emphasize that -- considering the alternative
frequencies suggested by the power spectrum and the limited quality of 
the radial velocity measurements -- the value of the orbital period chosen
here still requires confirmation. However, it appears that the current
observations are not compatible with the orbital period 
of 7\hochpunkt{h}59 
suggested by Patterson, marked by an arrow in the upper frame
of Fig.~\ref{hqmon-rv}.

\subsection{Photometry}
\label{HQ Mon Photometry}

\begin{figure}
   \parbox[]{0.1cm}{\epsfxsize=14cm\epsfbox{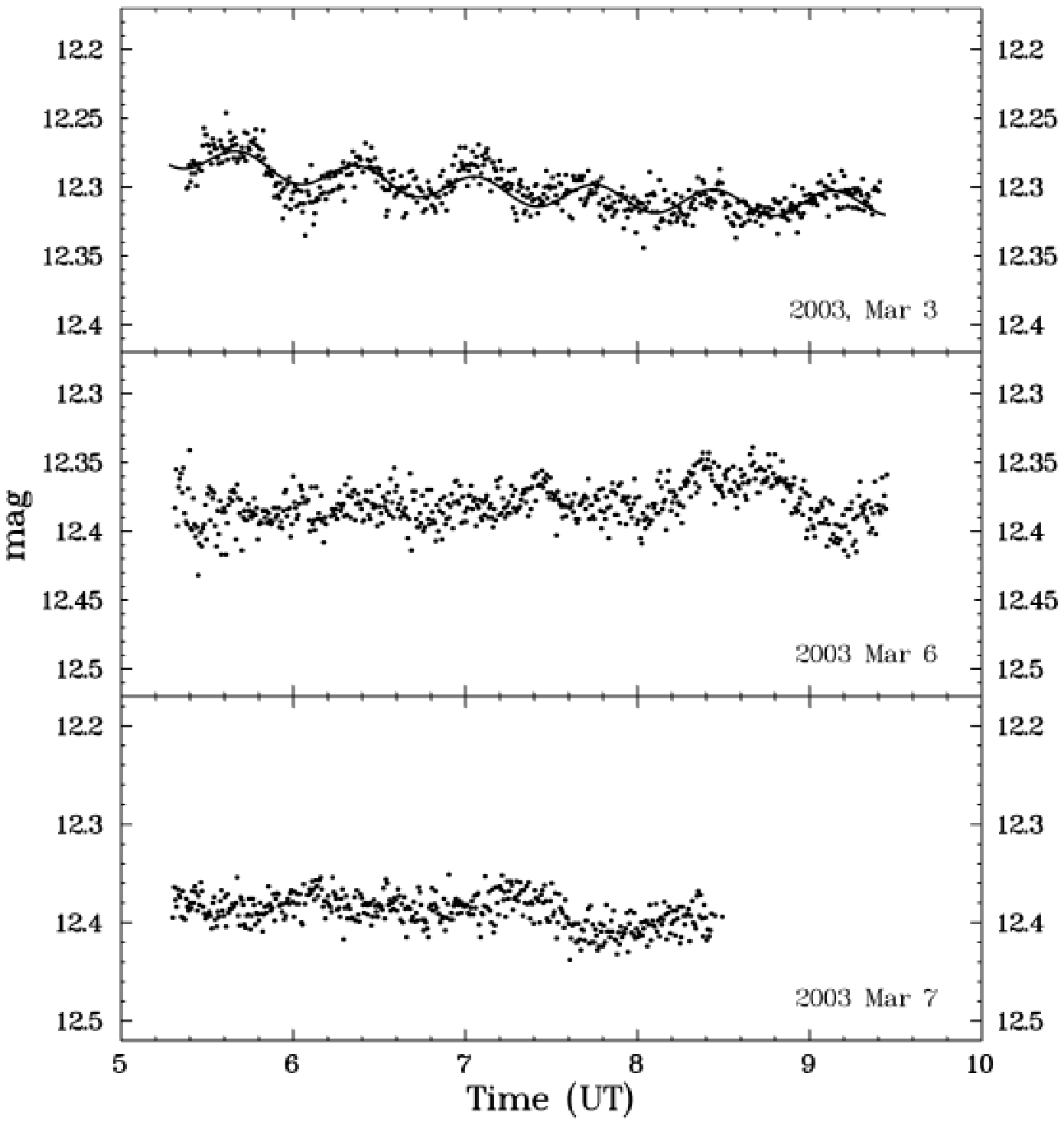}}
      \caption[]{Light curves with $\sim$$25^{\raisebox{.3ex}{\scriptsize s}}$
                 time resolution from the AAVSO data base. The photometric
                 band is not specified. The solid line in the upper frame
                 represents a formal two component sine fit to the data (see
                 text for details).}
\label{hqmon-aavso}
\end{figure}

\begin{figure}
   \parbox[]{0.1cm}{\epsfxsize=14cm\epsfbox{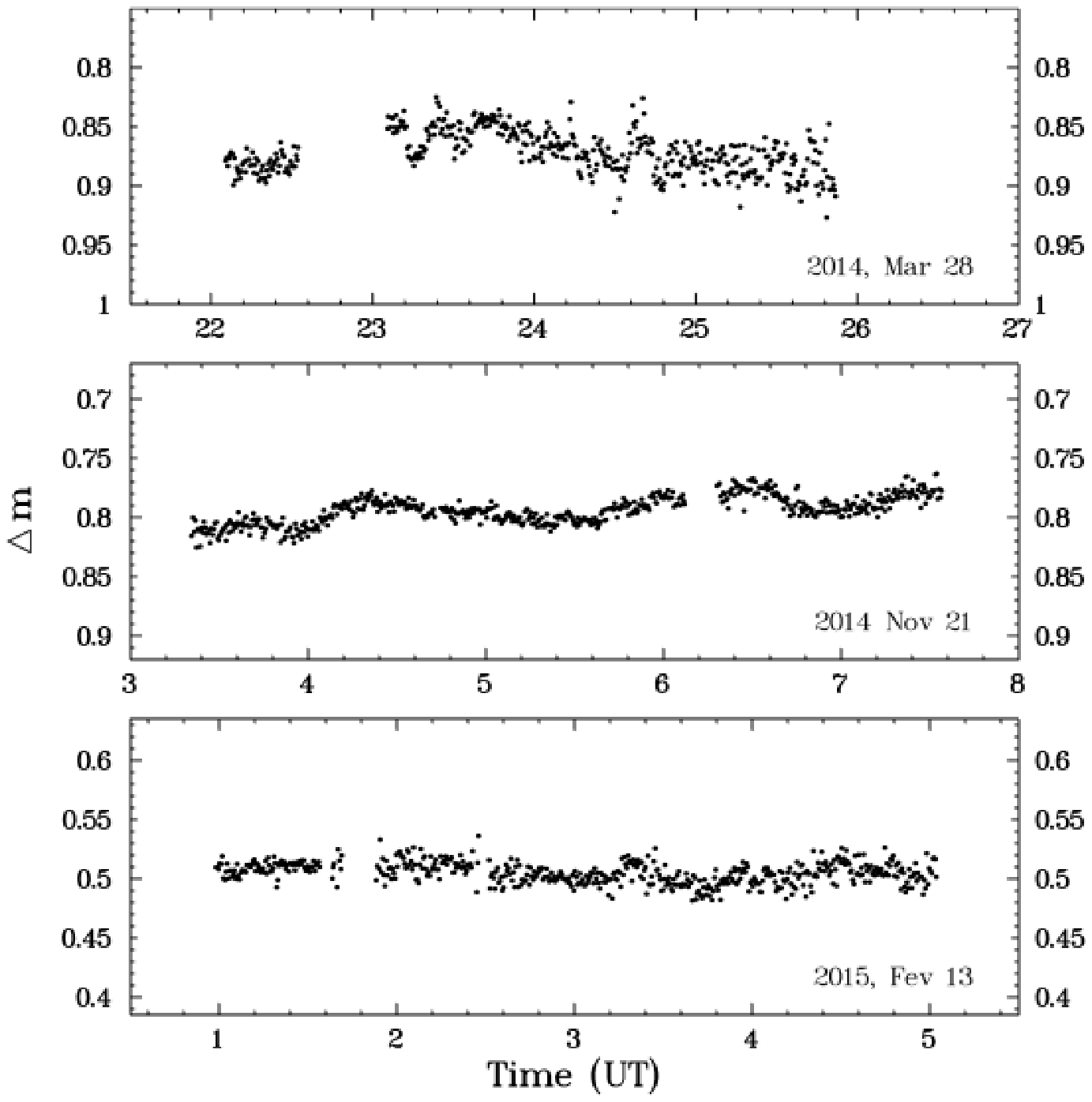}}
      \caption[]{Time resolved differential light curve of HQ Mon 
                 in three nights in 2014 and 2015.}
\label{hqmon-lightc}
\end{figure}

While no time resolved photometry of HQ~Mon has ever been published the AAVSO
data base contains light measurements performed with a cadence of 
$\sim$25\hoch{s} in three nights in 2003 March.
These are reproduced in Fig.~\ref{hqmon-aavso} in order to be compared with
the new data of 2014 and 2015, presented in this paper 
(Fig.~\ref{hqmon-lightc}), binned to the
same time resolution as the AAVSO data. To facilitate comparison the time
and magnitude scales are the same in both figures.

HQ~Mon exhibits night-to-night variations which can reach several tenths of
a magnitude. This is compatible with the historical records and with the
long-term AAVSO light curve (not shown) and is quite normal for cataclysmic
variables. Superposed on these long-term light modulation is low-scale
flickering with amplitudes of the order of 0\hochpunkt{m}03 -- 
0\hochpunkt{m}04\footnote{The larger scatter in the light curve of 
2014, March 28
(top panel of Fig.~\ref{hqmon-lightc}), in particular near the end, is partly 
due to the extension of the observation to quite high air masses.}, even 
smaller than observed in LS~IV~-08\hoch{o}~3 and the average UX~UMa stars (see 
Sect.~\ref{LS IV -08 3 Photometry}) but similar to the median flickering 
amplitude of 0\hochpunkt{m}035 observed in the novalike variable  
IX~Vel (calculated from the results of Beckemper 1995).
Considering that HQ~Mon does not exhibit dwarf nova outbursts and has
also never been seen in the long-term AAVSO light curve to go into a low 
state it appears therefore most appropriate to classify the system as a
{\em bona fide} UX~UMa variable. This
is in line with the spectral appearance i.e.\ hydrogen emission superposed on
broad absorption lines and He~II $\lambda$4686 in emission (Wahlgren et al.\
1985).

The light curve of 2003, March 3 (upper frame of Fig.~\ref{hqmon-aavso})
exhibits an intriguing pattern of apparently regular oscillations with 
decreasing amplitude superposed on a gradual fading. To parameterize it, two 
sine curves were fit to the data (solid line in the figure), one with a long 
period ($\sim 54$\hoch{h}) accounting for
the slow variations\footnote{This is only a parameterization! We neither 
claim that the slow variations are sinusoidal nor that they are periodic.}, 
and another with a short period following
approximately the oscillations. The latter turns out to be about 41\hoch{m}.
Based on this light curve alone one might suspect an intermediate polar
nature for HQ~Mon, the 41\hoch{m} oscillation reflecting the rotation period
of the white dwarf. However, since this pattern of variability does not repeat
itself in other nights, this might well be an over interpretation of the data.

\section{ST Cha}
\label{ST Cha}

The classification history of ST~Cha, discovered originally by 
Luyten (1934), was recently summarized by Simonsen et al.\ (2014). 
First considered to be an irregular variable or some kind of variable young 
star it entered the catalogues of CVs based on the suggestion of 
Cieslinski et al.\ (1998) who proposed a dwarf nova classification. This 
notion is founded on the spectrum of ST~Cha 
which shows a strong blue continuum with broad and shallow
absorptions in the higher Balmer lines and H$\alpha$ in emission (but note 
that the blue and red parts of the spectrum were observed in different nights).
Moreover, the $UBVRI$-colours of ST~Cha, as measured by 
Cieslinski et al.\ (1997) are compatible with the colours normally 
encountered in CVs (Bruch \& Engel, 1994; Echevarr\'{\i}a, 1988).

\begin{figure}
   \parbox[]{0.1cm}{\epsfxsize=14cm\epsfbox{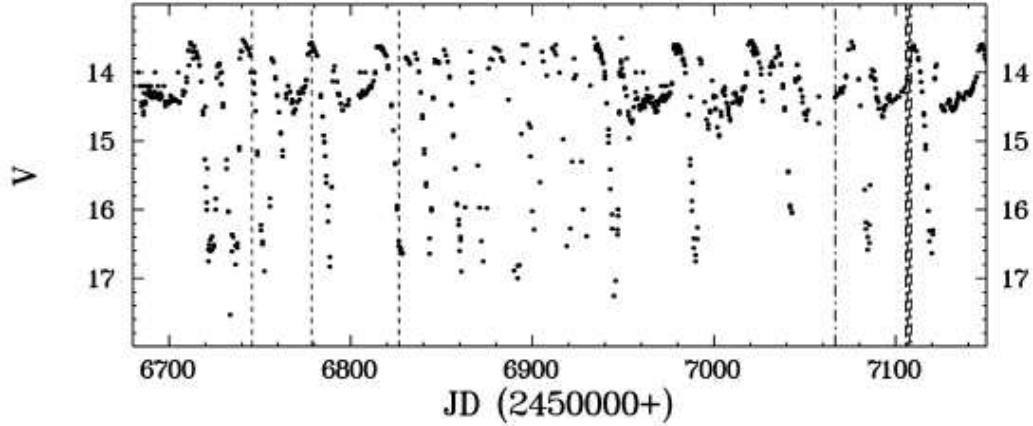}}
      \caption[]{Part of the AAVSO long-term light curve of 
                 of ST~Cha. The epochs of the present photometric
                 and spectroscopic observations are indicated by vertical 
                 dashed and dashed-dotted lines, respectively.}
\label{stcha-aavso}
\end{figure}

The nature of ST~Cha is discussed by Simonsenet al.\ (2014) based
on the long-term AAVSO light curve. In particular its more recent parts have
a dense coverage and therefore provide the most conclusive information. 
In Fig.~\ref{stcha-aavso} a 470 day section is shown, encompassing the dates 
of the presently discussed photometric and spectroscopic observations which 
are indicated by the vertical dashed and dashed-dotted lines, respectively. 
Most of the data points are CCD $V$ magnitudes with some visual magnitudes 
interspersed. The light curve reveals outbursts in rapid succession with only
quite short excursions to a faint (minimum) state, sometimes interrupted by
standstills at a level slightly below maximum magnitude. This behaviour is
the hallmark of Z~Cam stars, and consequently Simonsen et al.\ (2014) classify
ST~Cha as such. 

\begin{figure}
   \parbox[]{0.1cm}{\epsfxsize=14cm\epsfbox{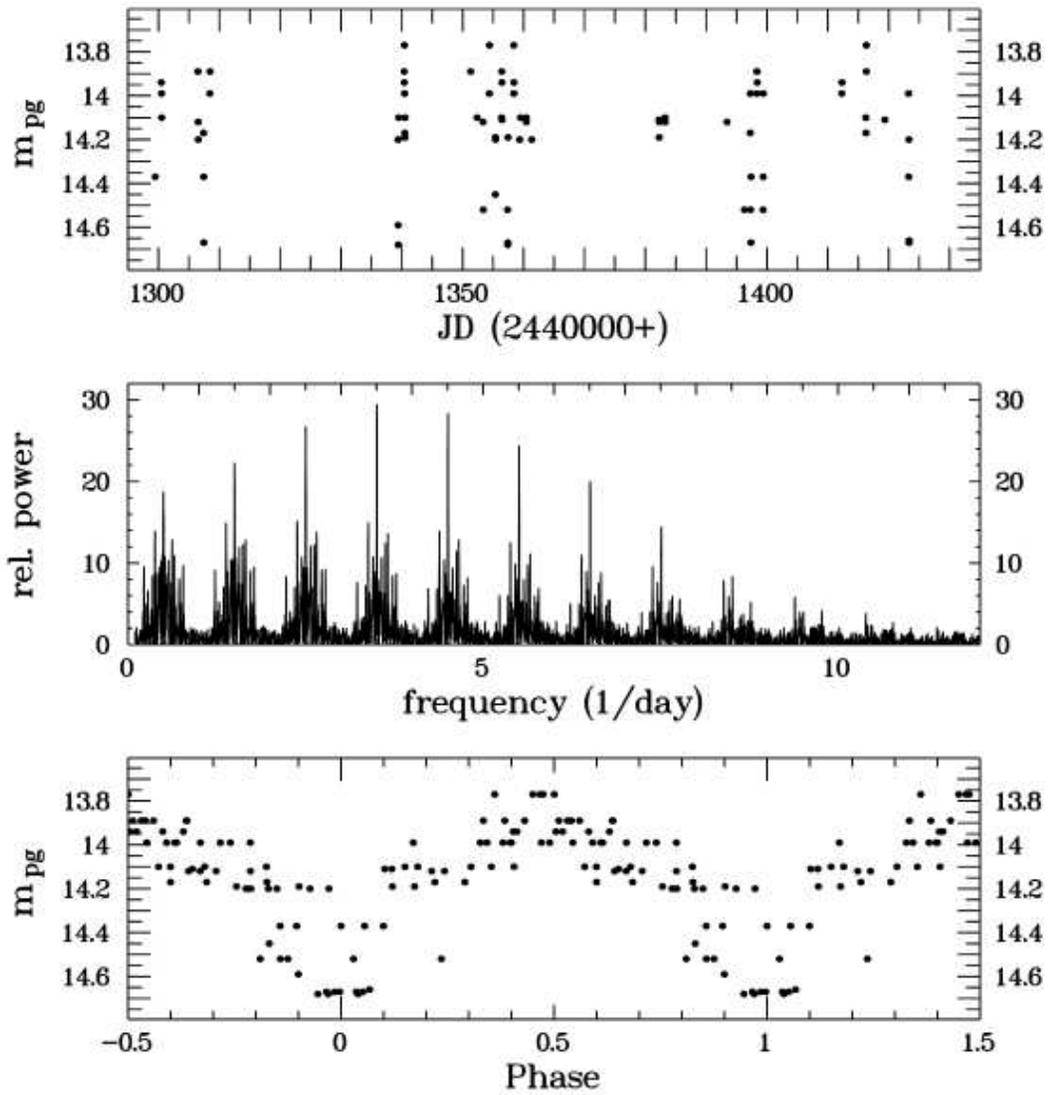}}
      \caption[]{{\em Upper frame:} Long-term photographic light curve of 
                 ST~Cha observed by Mauder \& Sosna (1975).
                 {\em Middle frame:} Lomb-Scargle periodogram of the light 
                 curve.
                 {\em Lower frame:} The light curve folded on the 
                 period 0.28535 days).}
\label{stcha-longterm}
\end{figure}

Cieslinski et al.\ (1998) mention that ST~Cha possibly exhibits
eclipses with a period of 6.8 hours (0.283 days) or 9.6 hours (0.4 days). 
This is based on a light curve published in tabular form by 
Mauder \& Sosna (1975), measured on 91 photographic plates distributed 
(inhomogeneously) over a time interval of 124 days in 1971-72. To have a 
closer look at this issue, we reproduce a graph of their
light curve in Fig.~\ref{stcha-longterm} (upper frame). The mean photometric 
magnitude is $\sim$$14\hochpunkt{m}25$, and the total range of variability 
spans about $1\hoch{m}$.

Various period search tools were applied to the light curve:
phase dispersion minimization (Stellingwerf 1978), analysis of 
variance (AoV; Schwarzenberg-Czerny 1989), and power spectrum analysis
following Lomb (1976) and Scargle (1982). While the latter 
is expected to be more sensitive to sinusoidal variations the other two 
methods depend less strongly on the waveform of a supposed periodic signal, 
but all of them yield compatible results. 

The Lomb-Scargle periodogram is shown in central frame of 
Fig.~\ref{stcha-longterm}.
It reveils a strong alias pattern with the highest peak at a frequency of 
3.505/day. This corresponds to a period of 0.2853 days or 6.847 hours, equal to
one of the periods suggested by Cieslinski et al.\ (1998). Their alternative
period corresponds to the frequency of the next strong alias peak at lower
frequencies in the power spectrum. 


Folding the data on the various periods it is difficult to decide which of the 
aliases is to be preferred. The lower frame of Fig.~\ref{stcha-longterm} shows
the folded light curve, using a period of 0.28535 days\footnote{No effort 
has been made to align the minimum in the folded light
curve to phase zero which is simply defined by the observing time of the first
data point in the light curve. Thus, the minimum coinciding with phase zero
is merely fortuitous.} (henceforth referred to as $P_{\rm MS}$). 
Considering the likely errors of photographic
photometry the resulting curve is satisfactory, but is is hardly the light
curve of an eclipsing cataclysmic variable.

As a caveat it must be mentioned that the alias frequencies in the periodograms
can all be explained as harmonics of one fundamental frequencies which 
corresponds exactly (within the resolution) to a period of 2 solar
days. While it
is not obvious what observational effect could cause a spurious period with
that value the very close coincidence of the fundamental period with a multiple
of one day may raise some concern as to the reality of the observed period. 

\subsection{Spectroscopy}
\label{ST Cha Spectroscopy}

\begin{figure}
   \parbox[]{0.1cm}{\epsfxsize=14cm\epsfbox{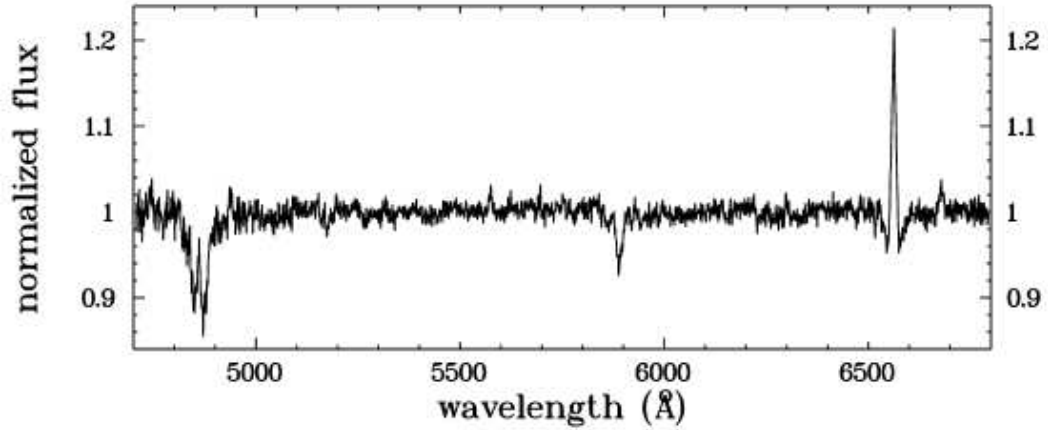}}
      \caption[]{Average continuum-normalized spectrum of ST~Cha}
\label{stcha-avspec}
\end{figure}

Spectra of ST~Cha were obtained in one night in 2015, February and in four
subsequent nights in 2015, March
(see Table~\ref{Journal of spectroscopic observations}). The AAVSO light
curve (Fig.~\ref{stcha-aavso}) shows that the February observations were taken
during a standstill, while in March ST~Cha was on the final rise from a
standstill to an outburst. The February data
are extremely noisy and will therefore not be regarded. During the four 
nights in March no significant changes of the spectrum were observed. Therefore,
the average spectrum, weighing the individual spectra by their mean count rates,
was calculated and is shown in Fig.~\ref{stcha-avspec}. 

The spectrum is dominated by a strong H$\alpha$ emission line located
within a shallow broad absorption component. H$\beta$ as a much fainter
emission component, while the absorption is significantly stronger than
in H$\alpha$. This behaviour is as expected for a dwarf nova in outburst
or standstill. The only other positively identified stellar line is a faint
emission of He~I at $\lambda$~6678~\AA. Another similarly strong structure
close to $\lambda$~5577~\AA\, is evidently caused by the telluric [O~I] 
auroral line not perfectly removed by the spectral extraction procedure.
ST~Cha being close to the maximum of an outburst makes it unlikely
that absorption features of the secondary star should be present in the
spectrum. This, together with the absence of any other late type absorption
feature suggests that the strong Na~D absorption lines seen just short ward
of $\lambda$~5900~\AA\, has an interstellar origin.

In order to measure the orbital period of ST~Cha, an attempt was made to
determine the radial velocity variations of the H$\alpha$ emission line,
using the cross correlation technique mentioned earlier.
Only the data of 2015, March 24 and
25 were considered because of the larger number of exposures in these nights
and their -- on average -- better signal-to-noise ratio as compared to the
spectra taken during the other nights. Even so, some individual spectra were
rejected because their H$\alpha$ line did not stand out sufficiently well from
the noise in the surrounding region. 

\begin{figure}
   \parbox[]{0.1cm}{\epsfxsize=14cm\epsfbox{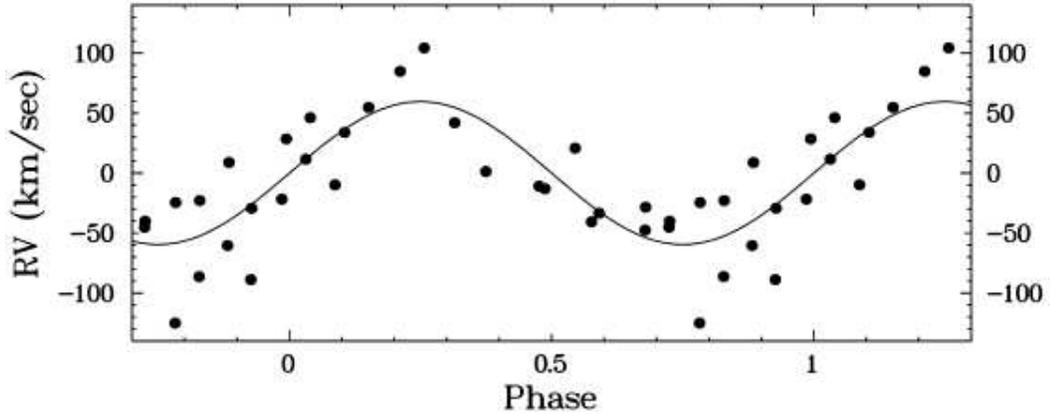}}
      \caption[]{Radial velocities of H$\alpha$ (dots) of ST~Cha observed on
                 2015, March 24 and 25, folded on the period $P = 0.229$~days
                 together with the best fit sine curve (solid line).}
\label{stcha-rvhalpha}
\end{figure}

The results were subjected to several period search routines. The AoV method,
the Lomb-Scargle periodogram, the Deeming (1975) power spectrum 
and a least squares sine fit all pointed at the same
period within a small range of 
0.006 days. The average value is 
$P_{\rm orb} = 0.229$ days which we will regard as the orbital period of 
ST Cha. The data folded on $P_{\rm orb}$ are shown together
with the best fit sine curve in Fig.~\ref{stcha-rvhalpha}. We cannot
exclude the possibility that the true period is a 1/day alias of this value.
However, folding the data on any one of these yields a less convincing RV 
curve. The radial velocity amplitude is 
60 $\pm$ 9 km/sec. 
The velocity zero point 
($\gamma$-velocity) is evidently determined by the peak of the average 
H$\alpha$ line and has no physical meaning. Therefore, $\gamma$ has been
subtracted from the curve in the figure. 

An error derived from the scatter of the period determined using the
different methods quoted above is unrealistically small because the
methods are not independent from each other. A more realistic period
error may be determined from the width of the peak corresponding to the
orbital frequency in the power spectra. The width (standard deviation)
of a Gaussian fit to 
this peak in the Deeming (1975) power spectrum corresponds to 
$\pm$0\hochpunkt{d}015.

\begin{figure}
   \parbox[]{0.1cm}{\epsfxsize=14cm\epsfbox{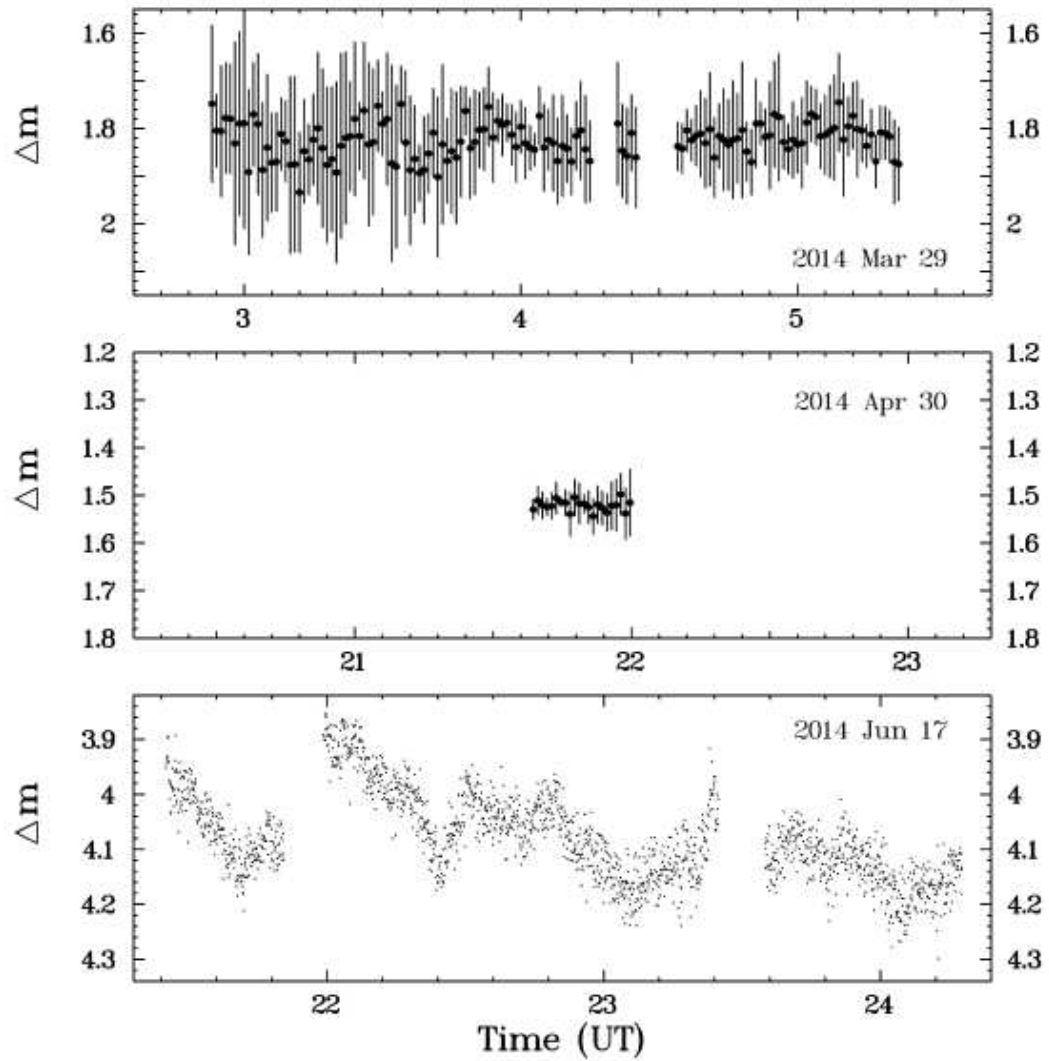}}
      \caption[]{Differential light curves of ST~Cha in three nights.
                 In the upper two frames the original time resolution
                 was degraded by binning the data into 1\hoch{m} intervals.
                 The vertical lines indicate the standard deviation of the
                 original data points around the bin average.}
\label{stcha-lightc}
\end{figure}

\subsection{Photometry}
\label{ST Cha Photometry}

We obtained time resolved differential photometry of ST~Cha in three nights in
2014. UCAC4~053-009413, taken from the Forth USNO CCD Astrograph Catalogue
(Zacharias et al.\ 2013) served as a comparison star. Its magnitude is
quoted by Zacharias et al.\ (2013) as $V=12\hochpunkt{m}317$.
The present light curves have an average differential magitude 
$\Delta$m of 1\hochpunkt{m}82 on March 29, 1\hochpunkt{m}52 on April 30 
and 4\hochpunkt{m}08 on June 17. Assuming that the effective wavelength 
of the current
white light observations corresponds roughly to $V$ the approximate visual
magnitude of ST~Cha during the three observing nights was thus 
14\hochpunkt{m}1, 13\hochpunkt{m}8 and 16\hochpunkt{m}4, respectively.
This is consistent with the magnitude of ST~Cha at the observing dates
in the long term light curve shown in Fig.~\ref{stcha-aavso}.

During the observations in March and April,
shown in the upper two frames of Fig.~\ref{stcha-lightc}, 
in particular on March 29, the observing conditions were far 
from ideal. Thus, the individual data points, observed with a time resolution 
of $\sim$5\hoch{s} show a large scatter. Therefore, 
for the purpose of visualization the data points were binned in intervals of
1\hoch{m} and the error bars shown in the figure  
represent the standard deviation of the individual data points within each
bin. 

On March 29, ST~Cha was starting a rapid decline from an outburst.
Because of the unfavourable weather conditions the time resolved light curve
is quite noisy even after data binning. The scatter of the binned data points
remains within the limits defined by the errors. Therefore, it is not
immediately clear if the observed variations are real -- due to flickering -- 
or if they are just caused by noise. An analysis of the correlation between
neighbouring data points, employing $R$ statistics, may help. This method,
first introduced by Baptista \& Steiner (1993) and later refined by 
Bruch (1999), basically determines whether neighbouring points in an 
ensemble of data scattered around zero have preferentially the same sign or
whether the sign changes statistically (as they should for a random data set
with zero mean).
Applying $R$ statistics to the residuals between the binned data points 
of 2015, March 29 and a straight line fitted by least squares to
these data, yields a probability of 0.9992 for correlations between the data
points. It may thus be concluded that the variations seen in the light curves
are not just accidental but that at least a part of them are real.

The light curve of April 30 is much shorter. It was observed right at
the maximum of an outburst following a standstill of ST~Cha. During this 
night the scatter of the individual data points is significantly
smaller\footnote{Note
that the magnitude scale in all panels of Fig.~\ref{stcha-lightc}
is the same.}. Again, $R$ statistics were applied to see if the 
residual scatter shows correlations or not. The probability that the data within
the observed $\sim$22\hoch{m} interval are indeed
correlated is low: 0.10. Thus, any flickering which might have been present
remained below the detection threshold.

The light curve of June 17 was obtained under much better atmospheric
conditions, but also in an different photometric state. As can be seen in
Fig.~\ref{stcha-aavso}, ST~Cha was close to the bottom of a short minimum 
between two outbursts. 
Considering the size of the telescope and the short exposure times the faint
magnitude explains the considerable noise in the light curve. Even so
considerable flickering activity is evident.

\subsection{The photometric period}
\label{The photometric period}

Having presented spectroscopic observation, the AAVSO light curve and 
additional 
time resolved photometry it is time to return to the question of the period 
$P_{\rm MS}$ derived from the observations of Mauder \& Sosna (1975). 

There are many reasons to suspect that there is something wrong with those
data. The foremost argument may be the contradiction with the spectroscopic
orbital period of ST~Cha. On the other hand, $P_{\rm MS}$ is based on 
photometric variations which may not reflect the orbital motion but some other
variability of the system. If so, it must have been transient because it
does not show up in more recent AAVSO data or in the present photometry.

Allowing for some observational scatter the total amplitude of the variations 
seen in the lower frame of Fig.~\ref{stcha-longterm} is of the order of
0\hochpunkt{m}8. The time between minimum and maximum is about 3\hochpunkt{h}4.
The mean photographic magnitude of 12\hochpunkt{m}2 
corresponds approximately to the average white light magnitude observed on
March 29. Since that light curve has a duration of about 
2\hochpunkt{h}5 any
variation of the type found in the data of Mauder \& Sosna (1975) should
have left a clear trace in the light curve, even considering the adverse
atmospheric conditions. The same is true for the even longer light curve
of June 17 which, however, was taken during a much fainter state.
The AAVSO light curve does not encompass the epoch of the Mauder \& Sosna (1975)
observations. It may not have the time resolution to show periodicities of 
the order of hours. However, if 
variations similar to those seen in Fig.~\ref{stcha-longterm} would have been
present during the longer standstills of ST~Cha 
(see Fig.~\ref{stcha-aavso}), they should have caused a considerably stronger
scatter in magnitude than actually observed. 

The origin of the strong
variations and the periodicity in the data of Mauder \& Sosna (1975)
therefore remains unclear. 
Having no convincing explanation we leave the question open.

\section{Conclusions}
\label{Conclusions}

With the intention to better characterize these system, we have presented
spectroscopic and photometric data of three bright cataclysmic variables.
Only for LS~IV~-08\hoch{o}~3 a reliably determined orbital period had 
already been published in the literature by Stark et al.\ (2008). The 
present data confirm many of their findings, permitted to improve the 
ephemeris of the system, and provided some additional information.
For the other two systems, HQ~Mon and ST~Cha, we publish for the first time
time-resolved spectroscopy and high time resolution 
photometry\footnote{noting that to our knowledge the high time resolution 
light curves of ST~Cha found on the AAVSO website have not otherwise been 
published.}. We find that previous unconfirmed values for the orbital
period of these systems, circulating in the literature, are erroneous and 
derived alternative values from radial velocity measurements. However, in
particular for HQ~Mon the proposed orbital period still requires independent
confirmation. 

\section*{Acknowledgements}

We gratefully acknowledge the use of the AAVSO data base which provided 
valuable supportive information for this study.

\section*{References}

\begin{description}

\item        Baptista, R., \& Steiner, J.E. 1993, A\&A 277, 311
\item        Beckemper, S. 1995, Statistische Untersuchungen zur St\"arke des
             Flickering in kataklysmischen Ver\"anderlichen,
             Diploma thesis, M\"unster
\item        Bruch, A. 1982, PASP, 94, 562
\item        Bruch, A. 1989, A\&AS, 78, 145
\item        Bruch, A. 1993, 
             MIRA: A Reference Guide (Astron.\ Inst.\ Univ.\ M\"unster
\item        Bruch, A. 1999, AJ, 117, 3031
\item        Bruch, A. 2016, New Astr., 46, 60
\item        Bruch, A., \& Engel, A. 1994, A\&AS, 104, 79
\item        Bruch, A., \& Schimpke, T. 1992, A\&AS, 93, 419
\item        Cieslinski, D., Jablonski, F.J., \& Steiner, J.E. 1997, 
             A\&AS 124, 55
\item        Cieslinski, D., Steiner, J.E., \& Jablonski, F.J. 1998, 
             A\&AS 131, 119
\item        Deeming, T.J. 1975, Ap\&SS, 39, 447
\item        Diaz, M. \& Hubeny, I. 1999, ApJ, 523, 786
\item        Eastman, J., Siverd, R., \& Gaudi, B.S. 2010, PASP, 122, 935
\item        Echevarr\'{\i}a, J. 1988, Rex.\ Mex.\ A\&A, 16, 37
\item        Hellier, C. 2001, Cataclysmic Variable Stars, Springer Verlag, 
             Berlin
\item        H{\o}g, E., Fabricius, C., Makarov, V.V., et al. 
             2000, A\&A, 355, L27
\item        Knigge, C. 2011, MNRAS, 373, 484 
\item        Knigge, C., Baraffe, I., \& Patterson, J.
             2011, ApJ Suppl., 194, 28 
\item        Lomb, N.R. 1976, ApSS, 39, 447
\item        Lloyd Evans, T. 1984, The Observatory, 104, 221
\item        Luyten, W.J. 1934, AN 253, 135
\item        Mauder, H., Sosna F.M. 1975, IBVS 1049
\item        Morgenroth, O. 1933, Astron.\ Nachr.\ 249, 385
\item        Munari, U., Zwitter, T., \& Bragaglia, A. 1997, A\&AS, 122,495
\item        Nassau, J.J., Stephenson, C.B. 1963, 
             Hamburger Sternw., Warner \& Swasey Obs.
\item        Ritter, H., Kolb, U. 2003, 
             A\&A, 404, 301
\item        Rosenfeld, A., Kak, C.A. 1982, Digital Picture Processing,
             Academic Press, New York, p.\ 353
\item        Samus, N.N., Durlevich, O.V., Kazarovetz, E.V., et al. 1975, 
             General Catalogue of Variable Stars, 
             http://www.sai.msu.su/gcvs/gcvs/
\item        Scargle, J.D. 1982, ApJ, 263, 853
\item        Schmidt-Kaler, Th. 1982, in: K.\ Schaifers, H.H.\ Voigt (eds.):
             Landolt B\"ornstein, Numerical Data and Functional Relationships in
             Science and Technology, New Series, Group VI, Vol.\ 2, Subvol.\ b,
             Springer Verlag, Heidelberg, p.\ 1
\item        Schneider, D.P., \& Young, P. 1980, ApJ, 238, 946
\item        Schreiber, M.R., Zorotovic, M., \& Wijnen, T.P.G. 2016
             MNRAS, 455, L16
\item        Schwarzenberg-Czerny, A. 1989, MNRAS, 241, 153
\item        Shafter, A.W. 1983, ApJ, 267, 222
\item        Shafter, A.W., Szkody, P., \& Thorstensen, J.R. 1986, ApJ, 308, 765
\item        Simonsen,M., Bohlsen, T., Hambsch, F.-J., \& Stubbings, R. 2014, 
             JAAVSO, 42
\item        Stark, M.A., Wade, R.A., Thorstenson, J.R., et al. 2008, 
             AJ, 135, 991
\item        Stellingwerf, R.F. 1978, ApJ, 224, 953
\item        van Bueren, H.G. 1950, Ann.\ Sterrew.\ Leiden,20, 201
\item        Wahlgren, G.M. 1992, AJ, 104, 1174
\item        Wahlgren, G.M., Wing, R.F., Kaitchuck, R.H., et al. 1985, 
             BAAS 17, 599
\item        Warner, B. 1995, Cataclysmic Variable Stars,
             Cambridge Astrophysics Series, Cambridge Univ.\ Press
\item        Zacharias, N., Finch, C.T., 
             Girard, T.M., et aql.\ 2013, AJ, 145, 44
\item        Zorotovic, M., Schreiber, M.R. \& G\"ansicke, B.T. 2011,
             A\&A, 536, A42
\item        Zwitter, T., \& Munari, U. 1994, A\&AS, 107, 503
\item        Zwitter, T., \& Munari, U. 1995, A\&AS, 115, 575
\item        Zwitter, T., \& Munari, U. 1996, A\&AS, 117, 44
\end{description}

\end{document}